\documentclass[11pt,twoside]{article}
\usepackage{asp2010}

\resetcounters

\begin{document}

\title{Chemical Tagging Hyades Supercluster as a consistency test of Stellar
Kinematic Groups.}
\author{H.M. Tabernero$^1$, D. Montes$^1$, and J.I. Gonz\'alez Hern\'andez$^{1,2}$
\affil{$^1$ Dpto. Astrof\'isica, Facultad de CC. F\'isicas,Universidad Complutense de Madrid, E-28040 Madrid,Spain.}
\affil{$^2$ Instituto de Astrof\'isica de Canarias, C/ Via
Lactea s/n, E-38200 La Laguna, Spain}
}

\begin{abstract}
Stellar Kinematic Groups are kinematical coherent groups of stars which may share a common origin. These       
groups spread through the Galaxy over time due to tidal effects caused by galactic rotation and disk heating, however the chemical information survives.
The aim of chemical tagging is to show that abundances of every element in the analysis must be homogeneous between members.
 We have studied the case of the Hyades Supercluster in order to compile a reliable list of members (FGK stars) based on chemical tagging information 
and spectroscopic age determinations of this supercluster. 
 This information has been derived from high-resolution echelle spectra obtained during our surveys of late-type stars. 
For a small subsample  of the Hyades Supercluster, stellar atmospheric parameters ($T_{\rm eff}$, $\log{g}$, $\xi$ and [Fe/H])
 have been determined using an automatic code which takes into account the sensibility of iron $EWs$ measured in the spectra.
 We have derived absolute abundances consistent with galactic abundance trends reported in previous studies. 
The chemical tagging method has been applied with a carefully differential 
abundance analysis of each candidate member of the Hyades Supercluster, using a well-known member of the Hyades cluster as reference.
A preliminary research has allowed us to find out which stars are members based on their differential abundances and spectroscopic ages.

\end{abstract}

\section{Sample selection}

The sample was selected using kinematical criteria in $U$, $V$ galactic velocities taking a dispersion of approximately 10 km/s around the core velocity of the group (Montes et al. 2001). \\
We have also taken additional candidates and spectroscopic information about some of these stars from L\'opez-Santiago et al. (2010), Mart\'inez-Arn\'aiz et al. (2010), and Maldonado et al. (2010). Some exoplanet-host star candidates are also taken from Montes et al. (2010).

\section{Observations}

Spectroscopic observations (see Fig. 2) were obtained at the 1.2 m Mercator Telescope at the \emph{ Observatorio del Roque de los Muchachos}  (La Palma, Spain) in January and May 2010 with HERMES (Raskin et al. 2010), a high resolution echelle spectrograph. 
The spectral resolution is 85000 and the wavelength range covers from $\lambda$3800~{\AA} to $\lambda$8750~{\AA} approximately. Our S/N ranges from 70 to 300 (120 on average) in the V band. A total of 61 stars have been observed. In this contributed paper only single main sequence stars (from F7 to K4) have been analyzed, being 42 in total.

\begin{figure}[!ht]
\plotone[scale=0.3]{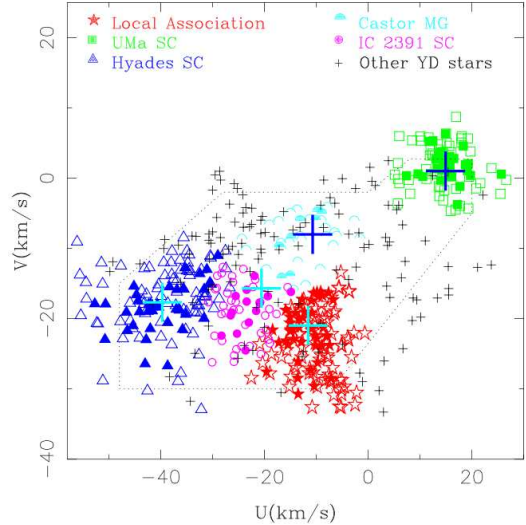}
\caption{$U$, $V$ velocities for late-type stars as possible members of different stellar kinematic groups (Montes et al. 2001). The blue triangles are the Hyades Supercluster candidates selected for this study filled symbols indicate stars that satisfy Eggen membership criteria  whereas open symbols represent other possible members.}
\end{figure}

\begin{figure}[!ht]
\plotone[bb= 40 83 500 657, clip=true, angle=90, scale=0.40]{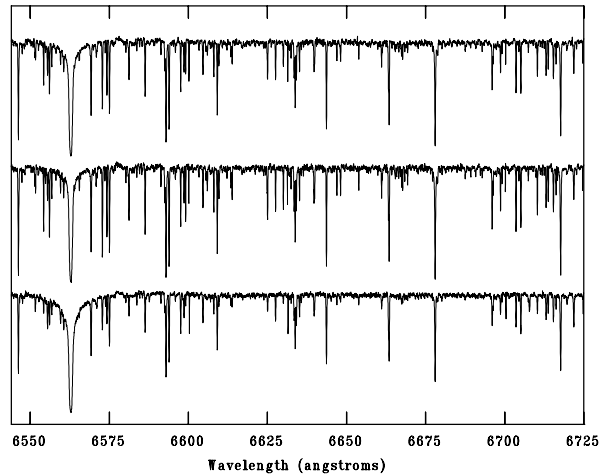}
\caption{High resolution spectra for some representative stars of our sample, from top to bottom: 
vB153, BZ Cet, HD 53532 (see Table 1). }
\end{figure}
 
\section{Stellar parameters}

Stellar atmospheric parameters ($T_{\rm eff}$, $\log{g}$, $\xi$ and [Fe/H]) have been determined with a own-developed code which iterates until the slopes of $\chi$ vs $\log{\epsilon(\textrm{Fe I})}$ and $\log{(EW / \lambda)}$ vs $\log{\epsilon(\textrm{Fe I})}$ where zero and imposing ionization equilibrium: $\log{\epsilon(\textrm{Fe I})} = \log{\epsilon(\textrm{Fe II})}$). \\
\\
Fig. 3 shows the $T_{\rm eff}$ and $\log{g}$ histograms for the stars analyzed. The obtained values for representative stars are given in  Table 1.

\begin{figure}[!ht]
\plotone[height=5cm, width=11cm, scale=0.7]{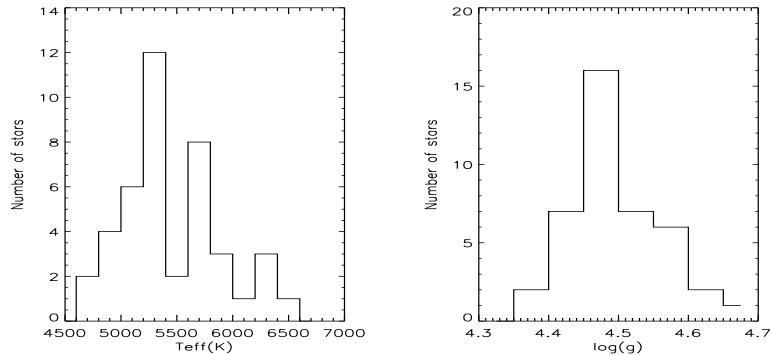}
\caption{$T_{\rm eff}$ and $\log{g}$ histograms of the sample }
\end{figure}
\section{Abundance determination}

Absolute abundances have been calculated using the equivalent width ($EW$) method in a line-by-line basis. Line lists were taken from Gonz\'alez Hern\'andez et al. (2010) and the $EW$ measured with ARES code (Sousa et al. 2007). Abundance analysis was carried out with the MOOG code (Sneden 1973) using the derived atmospheric parameters and  we used a HERMES spectrum of the asteroid Vesta taken with the same instrumental configuration as Solar reference. Our abundance trends seem to be consistent with the thin disk solar analogs (Gonz\'alez Hern\'andez et al. 2010) as shown in Fig. 4 (representative abundances are given in Table 2). 

\begin{figure}[!ht]
\plotone[scale=0.7]{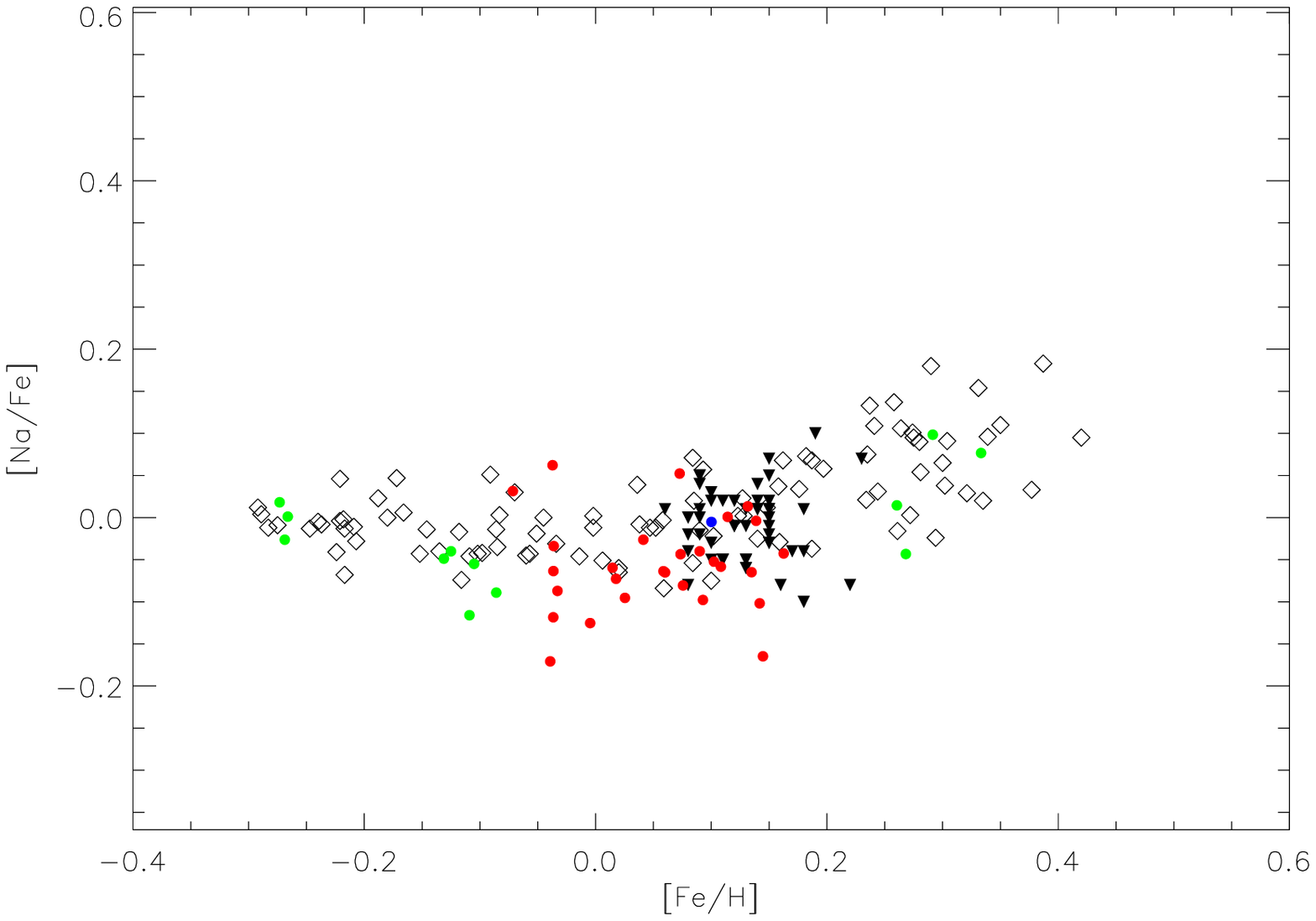}
\plotone[scale=0.7]{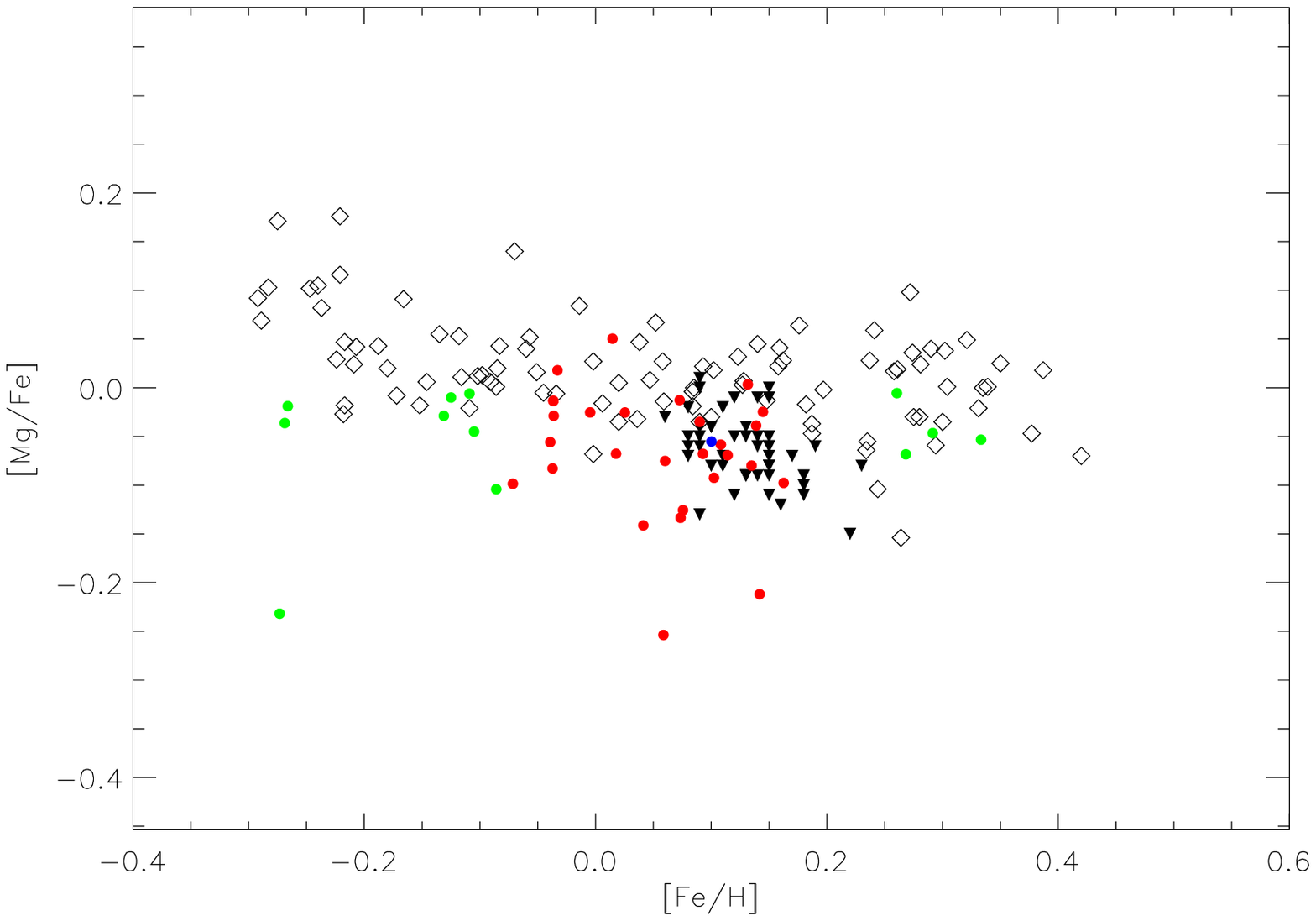}
\caption{[Na/Fe] and [Mg/H] vs [Fe/H]: open diamonds represent the thin disk data (Gonz\'alez Hern\'andez et al. 2010), black filled triangles represent Hyades cluster data (Paulson et al. 2003), red points are our stars compatible with Hyades Fe abundance provided by Paulson et al. (2003), and the green ones show no compatible stars. The BZ Cet Hyades cluster member star is marked with a blue point.}
\end{figure}

\begin{table}[!ht]
\begin{center}
\begin{tabular}{lccccccc}
\hline
 \textbf{Name} & \textbf{M$_{\bf V}$} & $\textbf{m}_{\bf V}$ & $ \textbf{T}_{\bf exp}$ & \textbf{S/N} &   \mbox{\boldmath${T}_{\rm eff}$} &     \mbox{\boldmath$\log{g}$} &  \mbox{\boldmath$\xi$} \\
           &     &                  &  {\small (s)} &  & {\small (K)}  &  & {\small (km/s)}  \\
              \hline
vB 153  &   5.96 &  8.90 & 3200 & 180& 5235 $\pm$ 36 & 4.45 $\pm$ 0.11 & 1.14 $\pm$ 0.06\\
 BZ Cet  &     6.13  &  7.98& 1200 & 130 & 5036 $\pm$ 45 & 4.39 $\pm$ 0.12 & 1.01 $\pm$ 0.11\\
 HD 53532 &  5.06 & 8.27 & 1400 & 110 &5677 $\pm$ 22 & 4.53 $\pm$ 0.07 & 1.14 $\pm$ 0.04\\
\hline
\end{tabular}
\caption{Absolute and apparent visual magnitudes, total exposure times, signal to noise ratio in the V band, stellar parameters  and typical  errors for the stars displayed in Fig.~1.  vB 153 is the reference star used in the differential analysis, BZ Cet is a Hyades confirmed member and HD 53532 is a Hyades Supercluster candidate star that satisfies chemical homogeneity.}
\end{center}
\end{table}

\section{Differential abundances}

Differential abundances $\Delta$[X/H] have been determined by comparison with a reference star known to be member of the Hyades cluster (vB 153) in a line-by-line basis (see Paulson et al. 2003 and De Silva et al. 2006). We have computed the differential abundances for the following elements: Fe, Na, Mg, Al, Si, Ca, Sc, Ti, V, Cr, Mn, Co, and Ni, the most representative of them are shown in Figs. 5 to 7.\\
  A first candidate selection within the sample has been determined by applying a $3 \sigma$ rejection for the Fe standard deviation in the Hyades cluster (0.05 dex, Paulson et al. 2003, see Fig. 5). In this subsample another $2.5 \sigma$ diagnostic has been applied in order to prove homogeneity in each element (see Figs. 6 and 7 ). 

\begin{table}[!ht]
\begin{center}
\begin{tabular}{lccccc}
\hline
 Name & [Fe/H] & [Na/H] & [Mg/H] & [Si/H] & [Ca/H]\\
              \hline
 vB 153        & 0.06  & -0.04 & -0.04  & 0.13  & 0.09\\
 BZ Cet        & 0.10  &  0.10 &  0.05  & 0.17  & 0.09\\
 HD 53532  &  0.09 & -0.01 &  0.03  & 0.09  & 0.06\\
\hline
\end{tabular}
\caption{Chemical abundances for the three stars explained in Table 1}
\end{center}
\end{table}

\begin{figure}[!ht]
\plotone[scale=0.7]{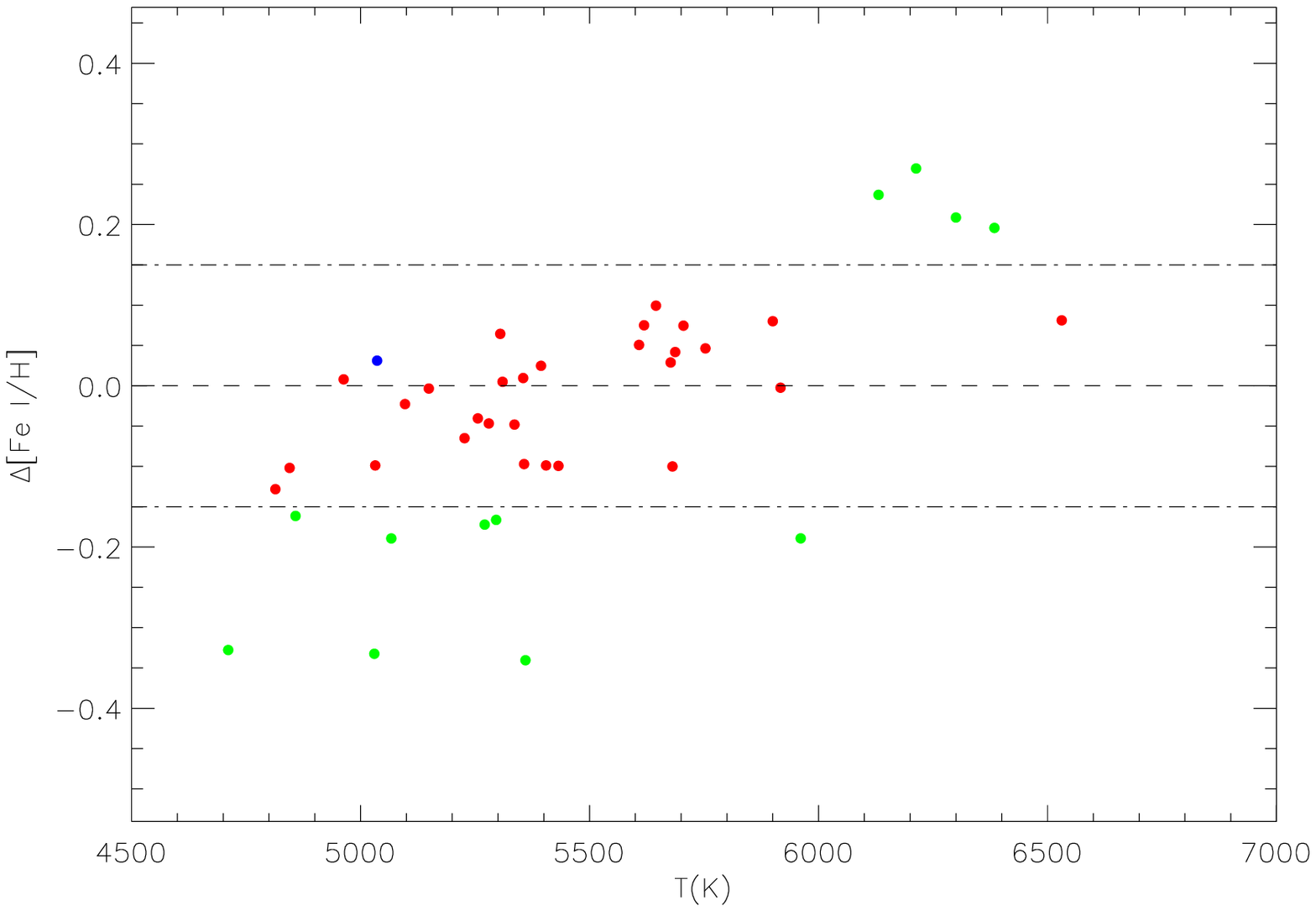}
\caption{$\Delta$[Fe/H] differential abundance vs $T_{\rm eff}$. Dashed-dotted lines represent $3\sigma$ level for the
Hyades. Red points are accepted as a preliminary selection of candidates, while green
ones are rejected. BZ Cet is denoted in blue. The Dashed line shows the mean abundance of the red points.}
\end{figure}

\begin{figure}[!ht]
\plotone[scale=0.8]{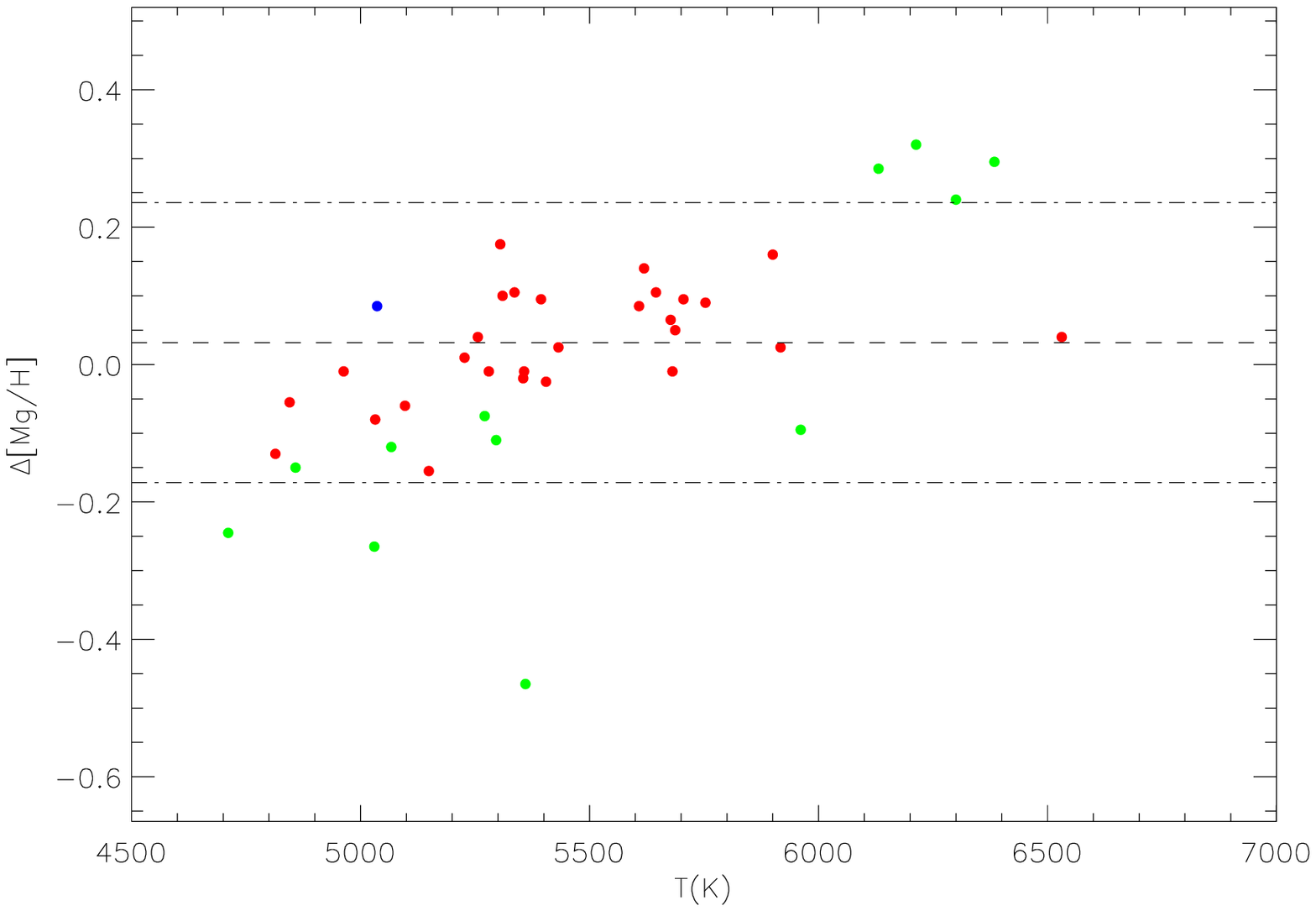}
\plotone[scale=0.8]{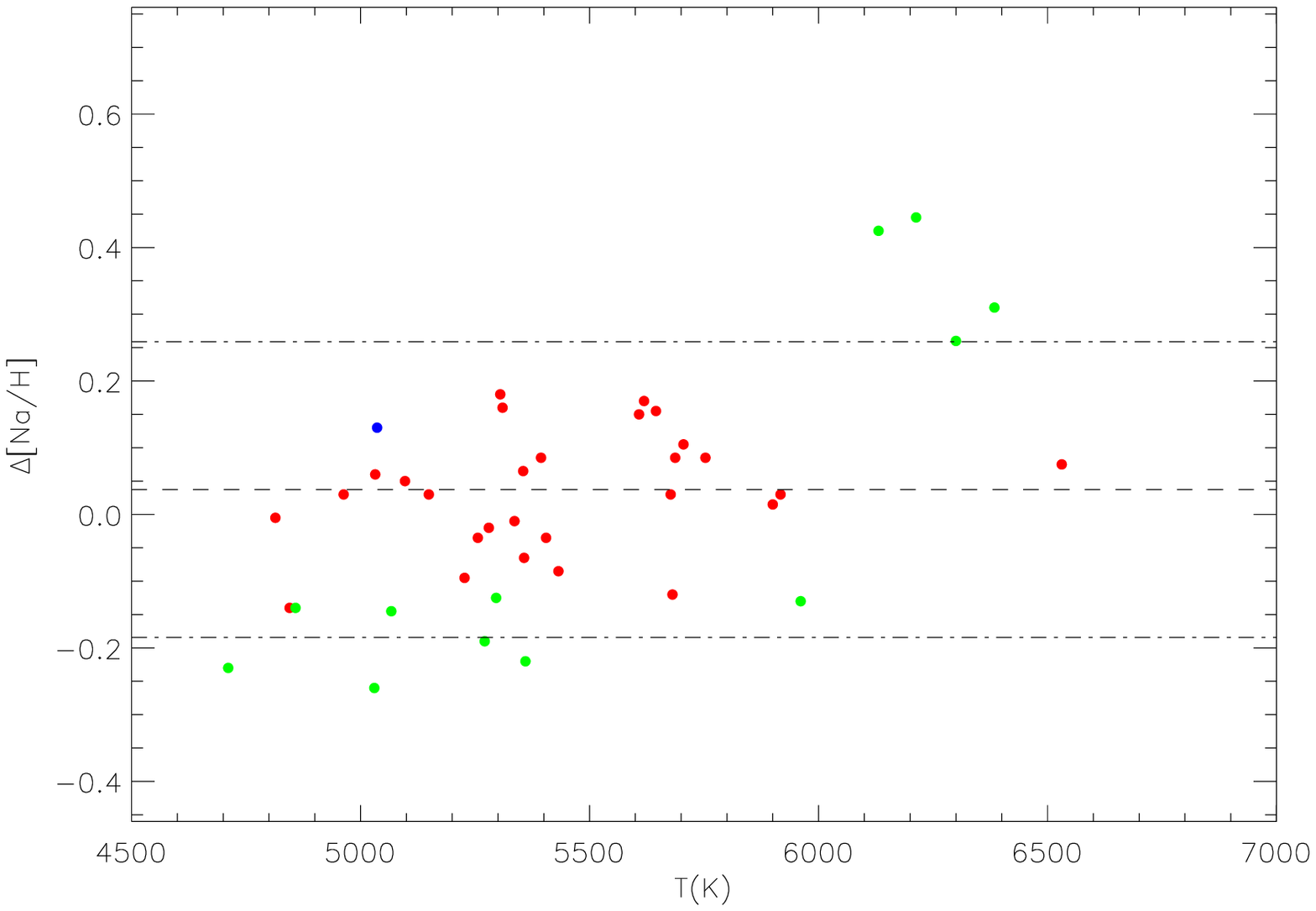}
\caption{$\Delta$[X/H] differential abundances (for Na and Mg) vs $T_{\rm eff}$. Dashed-dotted lines represent $2.5\sigma$ level for the
Hyades. Red points are accepted as a preliminary selection of candidates, while green
ones are rejected. BZ Cet is denoted in blue. Dashed line shows the mean abundance of the red points.}
\end{figure}
\begin{figure}[!ht]
\plotone[scale=0.8]{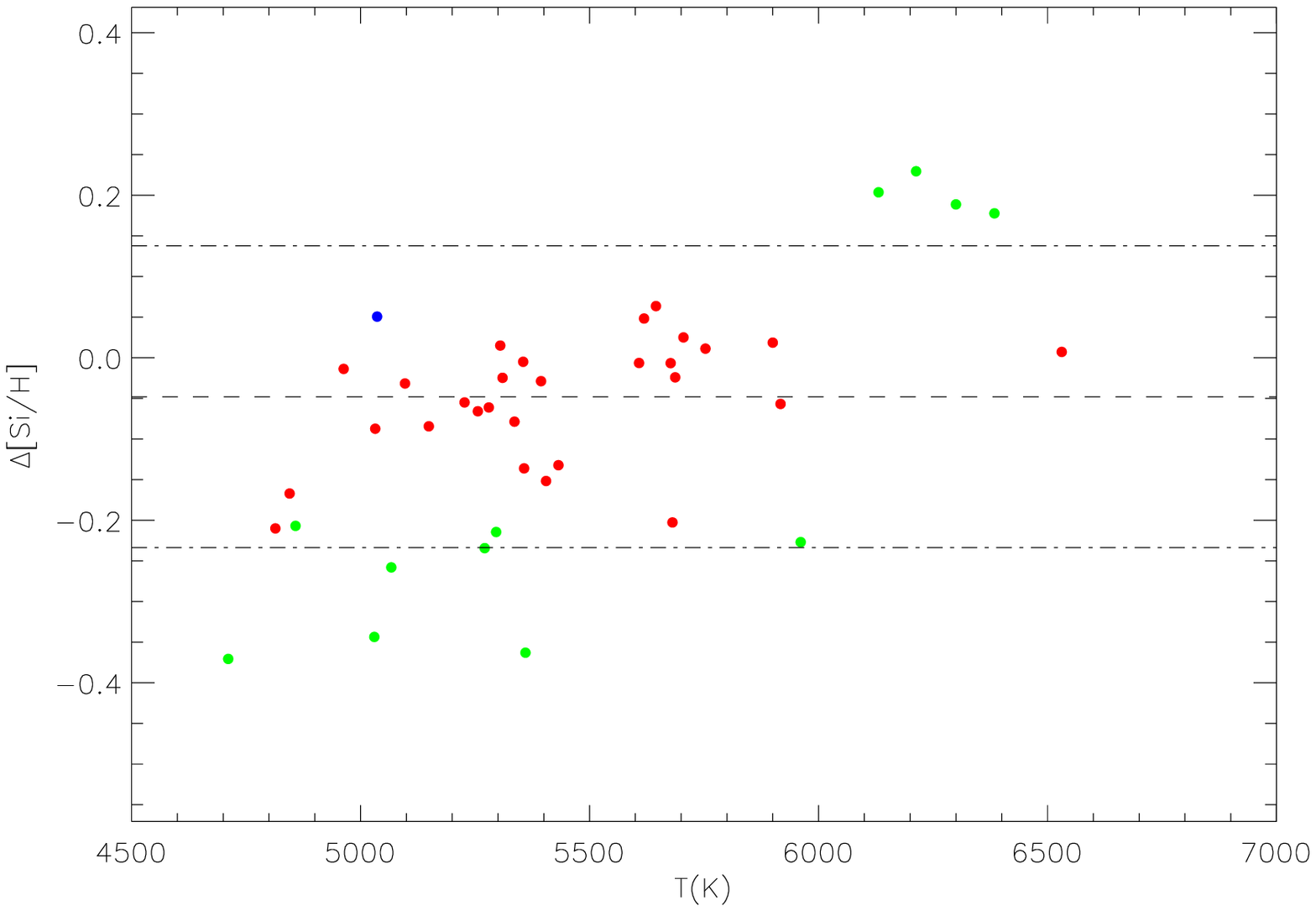}
\plotone[scale=0.8]{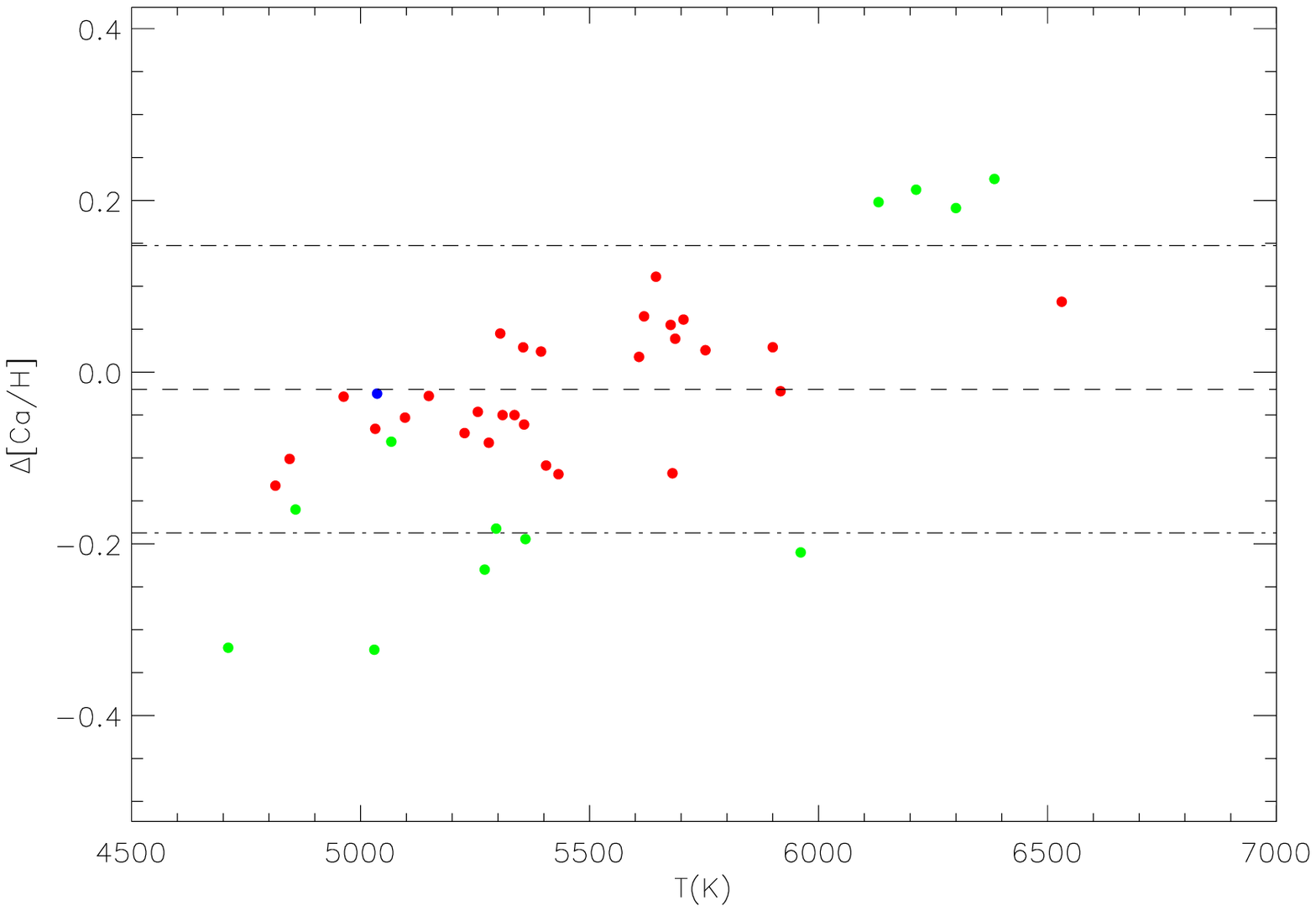}
\caption{Same as Fig.~6 for Ca and Si.}
\end{figure}

\section{Conclusions}

We have computed the stellar parameters and their uncertainties for 42 single main sequence Hyades Supercluster candidate stars, after that we have obtained the chemical abundances of 12 elements, and the differential abundances.
From the chemical tagging analysis we have found that 27 stars from the original sample are homogeneous in abundances for all the elements we have considered (a 64 \% of the sample), 3 stars fail to be homogeneous in one element.
 A more detailed analysis to check the consistency between the different age indicators and the chemical homogeneity is in progress.

\acknowledgements {This work was supported by the Universidad Complutense de Madrid (UCM), the Spanish Ministerio de Ciencia e Innovaci\'on (MCINN) under grant AYA2008-000695, and The Comunidad de Madrid under PRICIT project S2009/ESP-1496  (AstroMadrid).  Based on observations made with the Mercator Telescope, operated on the island of La Palma by the Flemish Community, at the Spanish Observatorio del Roque de los Muchachos of the Instituto de Astrof\'{\i}sica de Canarias.}

\bibliography{tabernero_h}

\nocite{*}
\end{document}